\documentclass[a4paper]{llncs}
\usepackage[utf8]{inputenc}
\usepackage{acronym}
\usepackage{pifont}
\usepackage{url}
\usepackage{balance}
\usepackage{algorithm2e}
\usepackage{xcolor}
\usepackage{graphicx}
\usepackage{amsmath}

\begin{document}
\mainmatter

\acrodef{ATC}{Air Traffic Control}
\acrodef{PSR}{Primary Surveillance Radar}
\acrodef{SSR}{Secondary Surveillance Radar}
\acrodef{FIS-B}{Flight Information Service - Broadcast}
\acrodef{TIS-B}{Traffic Information Service - Broadcast}
\acrodef{ADS-B}{Automatic Dependent Surveillance - Broadcast}
\acrodef{GPS}{Global Positioning System}
\acrodef{ES1090}{Extended Squitter - 1090 MHz}
\acrodef{UAT}{Universal Access Transceiver}
\acrodef{FAA}{Federal Avionics Administration}
\acrodef{FMS}{Flight Management System}
\acrodef{PPM}{Pulse-Position Modulation}
\acrodef{DF}{Downlink Format}
\acrodef{ICAO}{International Civil Aviation Organization}
\acrodef{RTCA}{Radio Technology Commission Aeronautics}
\acrodef{IBE}{Identity Based Encryption}
\acrodef{IBS}{Identity Based Signature}
\acrodef{PKI}{Public Key Infrastructure}
\acrodef{MAC}{Message Authentication Code}
\acrodef{HMAC}{Hashed Message Authentication Code}
\acrodef{COTS}{Commercial Off-The-Shelf}
\acrodef{ToA}{Time of Arrival}
\acrodef{TDoA}{Time Difference of Arrival}
\acrodef{CRC}{Cyclic Redundancy Check}
\acrodef{ABE}{Attribute Based Encryption}
\acrodef{KP-ABE}{Key-Policy - \ac{ABE}}
\acrodef{CP-ABE}{Ciphertext-Policy - \ac{ABE}}
\acrodef{AA}{Attribute Authority}
\acrodef{LSSS}{Linear Secret Sharing Scheme}
\acrodef{SIP}{System Identifying Party}
\acrodef{CA}{Certification Authority}
\acrodef{ACL}{Access Control List}
\acrodef{ACE}{Authentication and Authorization for Constrained Environments}
\acrodef{ABAC}{Attribute-Based Access Control}
\acrodef{PDP}{Policy Decision Point}
\acrodef{PEP}{Policy Enforcement Point}
\acrodef{PIP}{Policy Information Point}
\acrodef{PAP}{Policy Administration Point}
\acrodef{SDR}{Software Defined Radio}
\acrodef{CPR}{Compact Position Reporting}
\acrodef{ECDSA}{Elliptic Curve Digital Signature Algorithm}
\acrodef{TESLA}{Timed-Efficient Streamed Loss Tolerant Authentication}
\acrodef{DoS}{Denial of Service}

\newcommand\blfootnote[1]{%
  \begingroup
  \renewcommand\thefootnote{}\footnote{#1}%
  \addtocounter{footnote}{-1}%
  \endgroup
}

\newcommand{\TODO}{\textcolor{blue}{ \textbf{TODO}} }
\newcommand{\proto}{SOS}
\newcommand{\xmark}{\ding{56}}%
\newcommand{\cmark}{\ding{52}}%

\title{SOS - Securing Open Skies}
\titlerunning{SOS - Securing Open Skies}
\author{Savio Sciancalepore, Roberto Di Pietro}
\authorrunning{S. Sciancalepore and R. Di Pietro}
\institute{Division of Information and Computing Technology \protect\\ College of Science and Engineering, Hamad Bin Khalifa University \protect\\ Doha, Qatar \\ e-mail: ssciancalepore@hbku.edu.qa, rdipietro@hbku.edu.qa}

\graphicspath{{fig/}}

\maketitle

\begin{abstract}

    \blfootnote{The final version of this paper has been accepted to the SPACCS2018 conference. Please cite as: S.Sciancalepore, R.Di Pietro, "SOS - Securing Open Skies", Proceedings of the 11th International Conference on Security, Privacy and Anonymity in Computation, Communication and Storage (SPACCS 2018), Dec. 2018.}Automatic Dependent Surveillance - Broadcast (ADS-B) is the next generation communication technology selected for allowing commercial and military aircraft to deliver flight information to both ground base stations and other airplanes. Today, it is already on-board of 80\% of commercial aircraft, and it will become mandatory by the 2020 in the US and the EU. ADS-B has been designed without any security consideration --- messages are delivered wirelessly in clear text and they are not authenticated.

    In this paper we propose Securing Open Skies (SOS), a lightweight and standard-compliant framework for securing ADS-B technology wireless communications. \proto\ leverages the well-known $\mu$TESLA protocol, and includes some modifications necessary to deal with the severe bandwidth constraints of the ADS-B communication technology. In addition, \proto\ is resilient against message injection attacks, by recurring to majority voting techniques applied on central community servers.
    Overall, \proto\ emerges as a lightweight security solution, with a limited bandwidth overhead, that does not require any modification to the hardware already deployed. Further, \proto\  is standard compliant and able to reject active adversaries aiming at disrupting the correct functioning of the communication system.
    Finally, comparisons against state-of-the-art solutions do show the superior quality and viability of our solution.
\end{abstract}

\section{Introduction}
\label{sec:intro}

For years, the surveillance of air traffic has been performed through a combination of legacy radar technologies and human control \cite{Lim2017}. Communication systems such as the \ac{SSR} leverage on ground-based stations, that periodically interrogate transponders on-board of the aircraft to get information about the current status of the flight \cite{Strohmeier2015_survey}.

Starting from 2020, a new communication technology, namely \ac{ADS-B}, will become mandatory on all the commercial and military aircraft in the US and EU, by following specifications published by \ac{ICAO} and \ac{RTCA} \cite{ICAO_2014}. Anticipatory to the regulations, a few companies (e.g., Qatar Airways, American Airlines and British Airways) have already adopted the ADS-B standard.

\ac{ADS-B} uses the same frequency spectrum of the previous \ac{SSR} technology, but the communications are initiated by the aircraft, that periodically broadcasts messages reporting position, speed and other airplane-related information \cite{Strohmeier14_magazine}. 
On the one hand, \ac{ADS-B} provides a lot of advantages, both from the system perspective and from the costs side.  
On the other hand, it poses a lot of concerns regarding communication security. In fact, messages are delivered in clear text and without any inherent mechanism to guarantee their authenticity. This paves the way to a huge variety of threats, such as the one introduced by the capillary diffusion of cheap \acp{SDR}, able to inject custom-made packets in the air without requiring specific skills by operating entities \cite{Strohmeier2015_survey}.

Dealing with security issues in the context of avionic operations is a challenging task. In fact, avionic firms are often very slow to implement changes in their routines, due to business and regulatory concerns. In addition, the task is further complicated by both constraints in the communication bandwidth and the high message loss experienced on the single link due to obstacles and congestion \cite{Strohmeier2015_cps}. 
In the last years, with the approaching of the cited deadline, researchers from both academia and industry started formulating solutions to overcome these vulnerabilities. While a part of them focused on non-cryptographic security solutions, others still pushed for cryptography-based approaches. However, these latter contributions did not maintain compatibility with the latest standards, requiring substantial modifications to the message size, the available bandwidth, or the hardware to be used on-board of equipped aircraft (see Sec. \ref{sec:related} for a detailed overview).

\paragraph{Contributions} 
Our contributions are manifold. First, we propose Securing Open Skies (\proto), a standard-compliant framework integrating the well-known \ac{TESLA} protocol and allowing the verification of the authenticity of \ac{ADS-B} messages on a time-slot basis, without resorting to resource-demanding public-key cryptography solutions. Second, the integration is carried out in a standard compliant fashion.
Third, the framework allows for a joint processing of all the received packets on dedicated community servers, thus overcoming limitations due to the distributed nature of the network and the not negligible message loss on standalone receiving antennas. Moreover, \proto\ does not require hardware modification of the ADS-B receivers already deployed, thus being easy to integrate through a simple software update.
Finally, a thorough evaluation of \proto\ against competing solutions allows to establish its superior performance in terms of bandwidth overhead and provided security.

\paragraph{Roadmap} The paper is organized as follows: Sec. \ref{sec:related} reviews the recent literature on the topic; Sec. \ref{sec:prel_related} introduces the preliminary details about the \ac{ADS-B} technology, the \ac{TESLA} protocol and the adversary model; Sec. \ref{sec:proposal} provides the details of \proto, while Sec. \ref{sec:performance} analyzes the performance of the proposed solution and provides a comparison against state-of-the-art approaches, showing the superiority of our solution. Finally, Sec. \ref{sec:conclusion} tightens conclusions and draws future work.

\section{Related Work}
\label{sec:related}

The huge amount of work dealing with security in the context ADS-B technology can be divided in two main branches. From one side, grounding on the consideration that the scarce amount of bytes available in a ES1090 packet (see Sec. \ref{sec:preliminaries} for more details) does not allow for the inclusion of reliable cryptography solutions, many contributions focused on providing security services through additional system-level approaches.
To provide an example in this direction, the authors in \cite{Strohmeier2015_cps} propose a two-stage location verification scheme. During an offline stage it creates a fingerprint of a particular aircraft, leveraging both \ac{TDoA} values and deviations from nominal behavior. Then, in the online phase, it compares the received values with the fingerprint and evaluates the feasibility of the received data. In another work by the same authors \cite{Strohmeier2015_phy}, they propose an intruder detection algorithm based on the received signal strength, combining the measurements at the two antennas on board of an \ac{ADS-B} aircraft. Also, privacy issues are investigated in \cite{sciancalepore_percom2019}.

From the opposite side, other contributions still strive for cryptography based approaches, contextualizing their adoption in the severe constraints of the ADS-B technology. 
Authors in \cite{Baek2017} use a Staged \ac{IBE} (SIBE) scheme to provide confidentiality in \ac{ADS-B} communications. In their scheme, an aircraft uses the public key of a specific ground station to encrypt a message containing a random symmetric key. The ground station is the only entity able to decrypt the message with its private key, and then all subsequent communications use this new symmetric key. 
Even if the proposal is valuable, authors are converting a broadcast communication channel in a unicast communication channel, thus heavily modifying the logic and the functioning of the \ac{ADS-B} technology. Authors in \cite{Yang2017} propose a three-level Hierarchical \ac{IBS} (HIBS) scheme, in which each aircraft, associated to a given airlines recognized by a root authority (as \ac{ICAO} or EUROCONTROL) is able to sign its \ac{ADS-B} OUT messages by using keys generated according to its identity. Upon reception of a given signed message, a ground controller is able not only to identify the generating aircraft, but also its relationship with a given airline, approved by the root authority. 
However, being rooted on bilinear pairings, this scheme incurs a very high message fragmentation and overhead, thus being very hard to really be implemented in commercial aircrafts (more details will be provided in Sec. \ref{sec:comparison}).
In \cite{Kacem2015} and \cite{Kacem2016} the authors propose to use the \ac{HMAC} technique to assure integrity and authenticity of ADS-B messages. To reduce the message overhead of their solution, they split the cryptographic value between several concatenated messages, and verify the cryptographic validity of the \ac{HMAC} value only when all the portions are correctly received. However, the digests are computed over each single message, generating a very high communication overhead. In addition, they change the computation of the \ac{CRC} field, thus making their proposal not standard compliant. 

As for the adoption of the TESLA authentication scheme in the \ac{ADS-B} technology, only few previous contributions have discussed its feasibility. While \cite{Strohmeier2015_survey} briefly highlights potential benefits and drawbacks of such an approach, recent work \cite{Berthier2017} and \cite{Yang2017_globecom} delved into details, providing also an initial implementation of the solution using \ac{SDR}. However, these approaches are not standard compliant and they did not consider the constraints of the communication technology, neither with regards to the message size nor with respect to the severe bandwidth requirements highlighted in Sec. \ref{sec:preliminaries}. In addition, their integration in a complete security framework, as well as their interaction with a set of community receivers, is not considered.

To sum up, by considering both branches of the current literature discussed above, we highlight that cryptography-based solutions are the only possible way to secure the ADS-B system in a fully reliable fashion. However, a standard compliant solution that is able to integrate security services while maintaining the full compatibility with the standard and guaranteeing a tolerable overhead on the communication side is still missing. In this context, \proto\ emerges a standard-compatible approach, that integrates cryptography in the ADS-B communications by requiring a limited amount of additional packets to be exchanged on the wireless channel.   

\section{Preliminaries and Adversary Model}
\label{sec:prel_related}

\subsection{ADS-B in a nutshell}
\label{sec:preliminaries}

Despite its mandatory adoption on-board of commercial flights has been scheduled for the 2020, the \ac{ADS-B} technology was born in the late 1980s, in correspondence with the introduction of the satellite technology, and it was originally designed to work aside with legacy communication technologies such as \ac{PSR} and \ac{SSR} \cite{Kacem2016}.

\begin{figure}[htbp]
	\centering
	\includegraphics[width=.5\columnwidth]{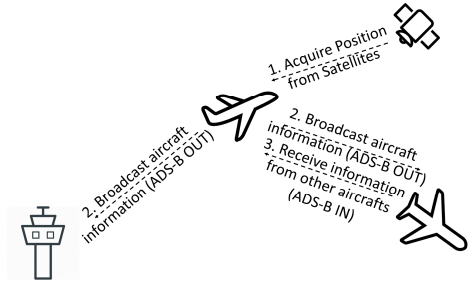}
	\caption{Overview of the ADS-B Communication Model.}
	\label{fig:scenario}
\end{figure}

The system has been designed to be \emph{Automatic}, given that it just needs to be turned on to work as intended, \emph{Dependant} because it requires dedicated operating airborne equipment, \emph{Surveillance}, because it is used as the primary surveillance method for controlling aircraft worldwide, and finally \emph{Broadcast}, due to the particular operational mode, in which the information is sent in broadcast \cite{Strohmeier2015_survey}.
The reference communication model is depicted in Fig. \ref{fig:scenario}.

An aircraft equipped with the ADS-B technology is able to obtain its position through satellites; then, it broadcasts its position via dedicated ADS-B messages.
The wireless operations can take place at two different frequencies: the 1090MHz frequency band, namely \ac{ES1090}, is used when the aircraft is above the height of 18,000 ft (about 5.5 km), while below this threshold the communications take place using the 978MHz frequency band, referred to as \ac{UAT}, to avoid further congestion on the ES1090 frequency band (due to the operation of previous technologies). In both cases, the dedicated channel bandwidth is 50kHz. The information delivered by the aircraft can be both received by \ac{ATC} ground stations, that can use them as a replacement or as a validation source for \ac{SSR}, or by other aircrafts.

The advantages deriving by the adoption of the \ac{ADS-B} technology are manifold. First, \ac{ADS-B} can improve pilots’ \emph{situation awareness}. In fact, pilots become able to receive traffic information about surrounding \ac{ADS-B} enabled aircraft, weather reports, and temporary flight restrictions. In addition, the cost of installing ADS-B ground stations is significantly cheaper with respect to installing and operating the \ac{PSR} and \ac{SSR} systems previously used. Moreover, \ac{ADS-B} provides better visibility to the aircraft with respect to legacy radar technologies, being able to guarantee an acceptable transmission range also in harsh regions (about 250 Nautical Miles, i.e., 450 km). 
At the data-link level, the \ac{ADS-B} message is encapsulated in Mode-S frames. As such, \ac{ADS-B} uses \ac{PPM} and the replies/broadcasts are encoded by a certain number of pulses, each pulse being 1 $\mu$s long \cite{Calvo2018}. 

From the system perspective, \ac{ADS-B} consists of two different subsystems, \emph{\ac{ADS-B} OUT} and \emph{\ac{ADS-B} IN}. \emph{\ac{ADS-B} OUT} is the service that allows the aircraft to periodically broadcasts information about the aircraft itself, such as identification information, current position, altitude, and speed, through a dedicated on-board transmitter. The \emph{\ac{ADS-B} IN} service, in parallel, allows for the reception of \ac{FIS-B}, \ac{TIS-B} data and other \ac{ADS-B} messages by the aircraft, as a result of a direct communication from nearby aircraft. 

\begin{figure}[htbp]
	\centering
	\includegraphics[width=.7\columnwidth]{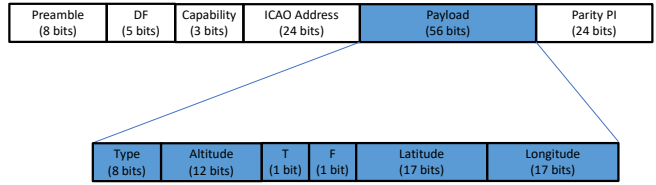}
	\caption{ADS-B ES1090 message format.}
	\label{fig:es1090pkt}
\end{figure}

\ac{UAT} and \ac{ES1090} have different payload requirements. The \ac{UAT} technology dedicates 272 bits (34 bytes) to the payload, while 36 and 112 bits are allocated for synchronization information (SYNC) and forward error correction parity information (FEC PARITY), respectively \cite{ICAO_2014}.
As for \ac{ES1090}, the structure  of the packet is showed in Fig. \ref{fig:es1090pkt}. 

While the preamble is used for synchronization purposes, the \ac{DF} field provides an indication of the transmission encoding, the Capability field is used to report the capability of an ADS-B transmitting installation that is based on a Mode-S transponder, the ICAO Address Field is reserved to the unique identification of the aircraft, while the Parity Information (PI) field provides error detection. A total number of 56 bits are reserved for the payload, where the Type field (8 bits) identifies the specific type of the payload message, the type T flag is used for synchronization purposes, the Subfield F flag indicates if the following position data are the even (0) or the odd (1) part of the message, while Altitude, Latitude and Longitude are reserved for data about the actual position of the aircraft. 

Finally, we highlight that the standard currently recommends (without forcing it) an overall maximum transmission rate of 6.2 messages per seconds, averaged over 60 seconds time interval.

\subsection{Security Considerations}
\label{sec:sec_cons}

The \ac{ADS-B} protocol does not include any security mechanism. Indeed, messages are transmitted in clear-text, allowing anyone equipped with a compatible receiver to decode their content and easily access to the information contained therein. This choice was done in the 80s to boost message availability. However, nowadays it is the cause of dreadful threats associated with the operation of the ADS-B technology. In fact, the wide availability of cheap \ac{COTS} \acp{SDR} opens the possibility to easily inject custom-made ADS-B messages on the wireless communication channel. Thus, it is very easy to perform a number of message injection attacks, including Aircraft Spoofing, Ghost Aircraft Injection/Flooding, Aircraft Disappearance, and Trajectory Modification, to name a few \cite{DiPietro2005}, \cite{Strohmeier2015_survey}.

However, the public availability of aircraft's data has the potential to strengthen the control on the avionic traffic and help establishing open initiatives to maintain the security of the sky navigation. In fact, the openness of the system inspired the rise of many collaborative networks, such as the \emph{OpenSky-Network} project \cite{Schafer2014}. OpenSky-Network is a community-based receiver network, which continuously collects \ac{ADS-B} data delivered from operational airplanes. In addition, OpenSky-Network makes data accessible to researchers worldwide for experimentation and testing.

As it will clearly emerge from the discussion in the following sections, the \proto\ protocol leverages a community-oriented approach on the receiver side, inspired by the presence of projects such as the OpenSky-Network. This allows the overall system to be inherently able to overcome limitations such as the potential loss of messages and the limited computational capabilities of single receiver antennas.

\subsection{Adversary Model}
\label{sec:adv_model}

In this work we assume a very powerful attacker, characterized by both passive and active features. The adversary is able to eavesdrop all the communications on the 1090 MHz frequency band, by assuming the use of \ac{COTS} devices such as a \ac{SDR} \cite{Tuttlebee2002}. Moreover, it is also able to inject fake messages over the wireless communication channel, by pretending to be a legitimate aircraft. This is indeed possible thanks to the presence of cheap \acp{SDR}, held at the ground level, able to forge fake messages and deliver them on the wireless communication channel. We also assume that the adversary, in order to stay stealthy, follows the constraints of the \ac{ADS-B} technology on the transmission rate: thus, it injects packets with a transmission rate within the limits imposed by the standard. Finally, we assume that the adversary is able to carry on the attack only for a reduced portion of the area covered by the flight, i.e., it is static and does not move with the aircraft.

\subsection{The TESLA protocol}
\label{sec:tesla}

The \acl{TESLA} (TESLA) protocol was initially proposed in \cite{Perrig2000} to authenticate media streams in a lightweight and time-efficient way, without resorting to resource-consuming public key cryptography solutions.
In \ac{TESLA} the time is divided in \emph{epochs}, with each epoch $i$ having a well-defined starting and ending time. It also assumes a loose synchronization between the communicating parties.
To provide authentication of broadcast messages, the entity that generates the messages is equipped with an initial secret, namely the \emph{root key}, shared only with a well-known authority, known to all the parties. At the boot-up of the system, the authority provides an initial key, namely \emph{key chain commit}, generated by hashing the root key a number $n$ of consecutive times. This element is shared on the communication channel and it is known to all the parties involved in the communication. A message, i.e., $m_i$, is authenticated by appending a \ac{HMAC} generated through a key $K_i$, obtained by hashing the initial \emph{key chain commit} exactly $n-i$ times.

The security of the scheme lies in the fact that the key used to generate the \ac{HMAC} in the epoch $i$ is not shared before the ending of the epoch itself. Thus, the receiving entities simply store the messages received in the slot, but they cannot verify them immediately (because of the lack of knowledge about the symmetric key). Only after a disclosure lag $d$ in epochs, the key is disclosed (in broadcast) on the communication channel and included in all the packets generated exactly $d$ epochs after, allowing the verification of all the messages delivered by the transmitting entity exactly $d$ epochs before. 
Note that the key disclosed by the transmitting entity is assumed to be genuine only if it allows, by $i$ consecutive hashing operations, to obtain exactly the \emph{key chain commit}. In this way, because of the one-way features of the hashing operation, the authenticity is guaranteed.
Despite its success and wide adoption, TESLA was not designed for severe constrained environments. To cope with this limitation, in $\mu$TESLA the key is not disclosed in each packet, but only once per epoch \cite{Perrig2002}. In addition, taking care of the constraints in the size of the memory of sensors, $\mu$TESLA also restricts the number of authenticated senders, thus limiting the memory footprint of the protocol. 

As it will emerge in the following sections, the proposed framework leverages the core logic of the $\mu$TESLA protocol, even if it provides further modifications necessary to deal with the limited payload size of \ac{ADS-B} messages. 

\section{The \proto\ framework}
\label{sec:proposal}

\subsection{Preliminary considerations}
\label{sec:prel_proto}

The system scenario assumed hereby involves the following actors:
\begin{itemize}
	\item \emph{Aircraft}. It is an \ac{ADS-B} equipped plane, emitting standard-compliant \ac{ADS-B} messages.
	\item \emph{Avionics Authority}. It is a super-parties authority, whose responsibility is to assign cryptography materials and unique addresses to operating aircraft. It is assumed to be online at least for a small amount of time during the operation of the aircraft. This role is the one natively assumed by ICAO and EUROCONTROL.
	\item \emph{Receiver Antennas}. They are a set of \ac{ADS-B} receivers, distributed over a large area, able to receive and successfully decode the messages delivered by the aircraft. In addition, they are supposed to forward the received messages to a remote server. This role is actually played by OpenSky Receiver Antennas.
	\item \emph{Community}. It represents a set of general-purpose servers that receive messages from the distributed antennas and provide additional computing intelligence to validate their authenticity and web-oriented services. This role is actually played by the OpenSky-network project. 
\end{itemize}

In the following we assume that the legitimate ADS-B-equipped aircraft has already taken off from an airport, and it has exceeded the altitude of 5,500 meters. Thus, it switches from \ac{UAT} to \ac{ES1090} mode, and starts emitting standard ADS-B messages. The set of wireless receivers in its communication range, equipped with ADS-B decoders, are able to detect and decode the messages. Next, they deliver all the messages to the servers community. 
The aim of the \proto\ framework is to provide authentication of the messages that have been effectively transmitted by the transmitting plane.

We also assume that the receivers and the transmitter are loosely synchronized with a common clock source, such as the UTC or the GPS system.
In addition, the time is divided in time-slots of a given duration $d_i$. Assuming $t_0$ is the time of the boot-up of the aircraft, the time-slot $t_i$ will trigger at the absolute value $t_i = t_0 + \sum_{j=0}^{i-1} d_j$.
Finally, without loss of generality, we assume that legitimate aircraft deliver \ac{ADS-B} messages at a constant rate of 6 packet/seconds, in line with constraints defined by the standard for the maximum allowed transmission rate for each aircraft \cite{ICAO_2014}.

\subsection{Extending the ADS-B protocol}
\label{sec:new_messages}

\proto\ provides the authentication of broadcast messages by leveraging symmetric cryptography techniques, without modifying the legacy structure of \ac{ADS-B} messages.
To this aim, we extend the \ac{ADS-B} technology, while pursuing standard-compliance, by adding new type of messages dedicated to the delivering of cryptography elements.

\begin{figure}[htbp]
	\centering
	\includegraphics[width=.7\columnwidth]{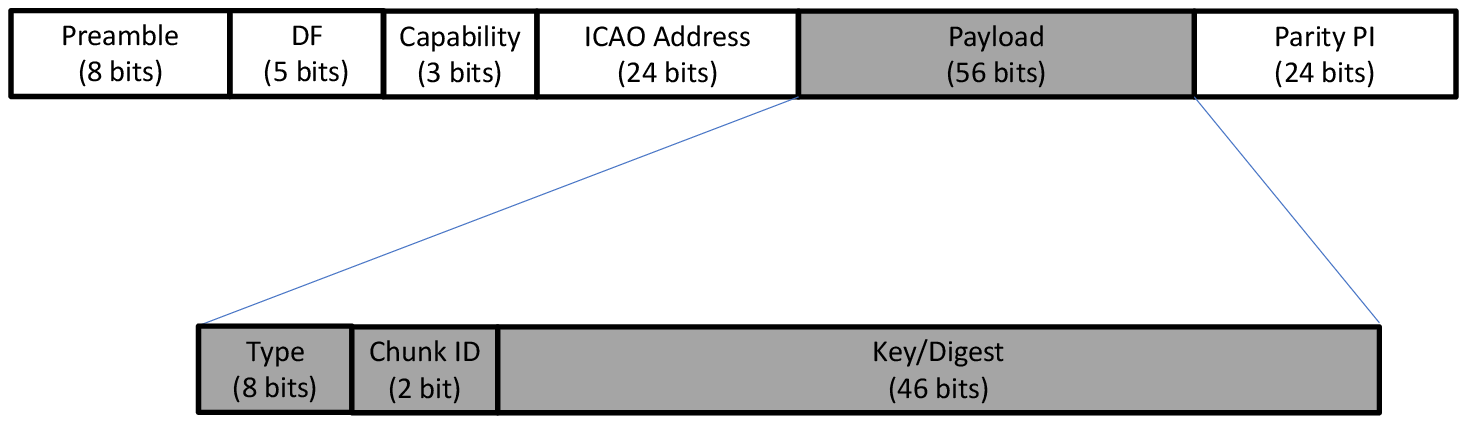}
	\caption{The content of verification packets transmitted by adopting \proto.}
	\label{fig:prop_pkt_2}
\end{figure}

The security messages are included in the \ac{ADS-B} packet as a part of the payload, leveraging the sub-field \emph{Type} of the message and specific values whose meaning is reserved for future use by the standard. A sample picture of the structure of security packets is provided in Fig. \ref{fig:prop_pkt_2}. 

The following two verification messages are defined:
\begin{itemize}
	\item Verification Digest, $Type=25$. This message is used to allow for the transmission of a message digest at the end of a slot by an aircraft. 
	\item Verification Key, $Type=32$. This message is used to transmit a verification key used in the previous slot, allowing the verification of the full batch of messages.
\end{itemize}

When the \emph{Type} field in the payload is either 25 or 32, the following part of the payload includes the following sub-fields:
\begin{itemize}
	\item Chunk ID (2 bits). It specifies the unique identifier of the portion of the following content included in this message. 
	\item Content (46 bits). It contains the effective payload of the verification message. In case the \emph{Type} field was 25, it contains the portion of the digest. Otherwise, in case the \emph{Type} field was 32, it includes the specified segment of the verification key for the previous slot.
\end{itemize}

\subsection{Details of the \proto\ framework}
\label{sec:protocol_details}

\proto\ provides messages authentication leveraging delayed hash chains. While it is inspired by the $\mu$TESLA protocol proposed in \cite{Perrig2002}, it presents several modifications made in order to adapt the protocol to the more severe constraints of the \ac{ADS-B} technology.

Overall, an \ac{ADS-B} receiver system that runs the \proto\ framework can work in two modes:
\begin{itemize}
    \item \emph{Unsecured Mode}: The receiver does not verify the authenticity of packets received through the receiver antenna. Thus, as soon as the packet is correctly decoded, the information are processed. The new \ac{ADS-B} messages having the Payload Type Field equal to 25 or 32 are simply discarded.
    \item \emph{Secured Mode}. As soon as the messages are decoded, they are buffered until the related verification digest and verification code are received. Only if the pool of messages is verified through the procedure described below, the information contained therein are further processed.
\end{itemize}

\begin{figure}[htbp]
	\centering
	\includegraphics[width=.6\columnwidth]{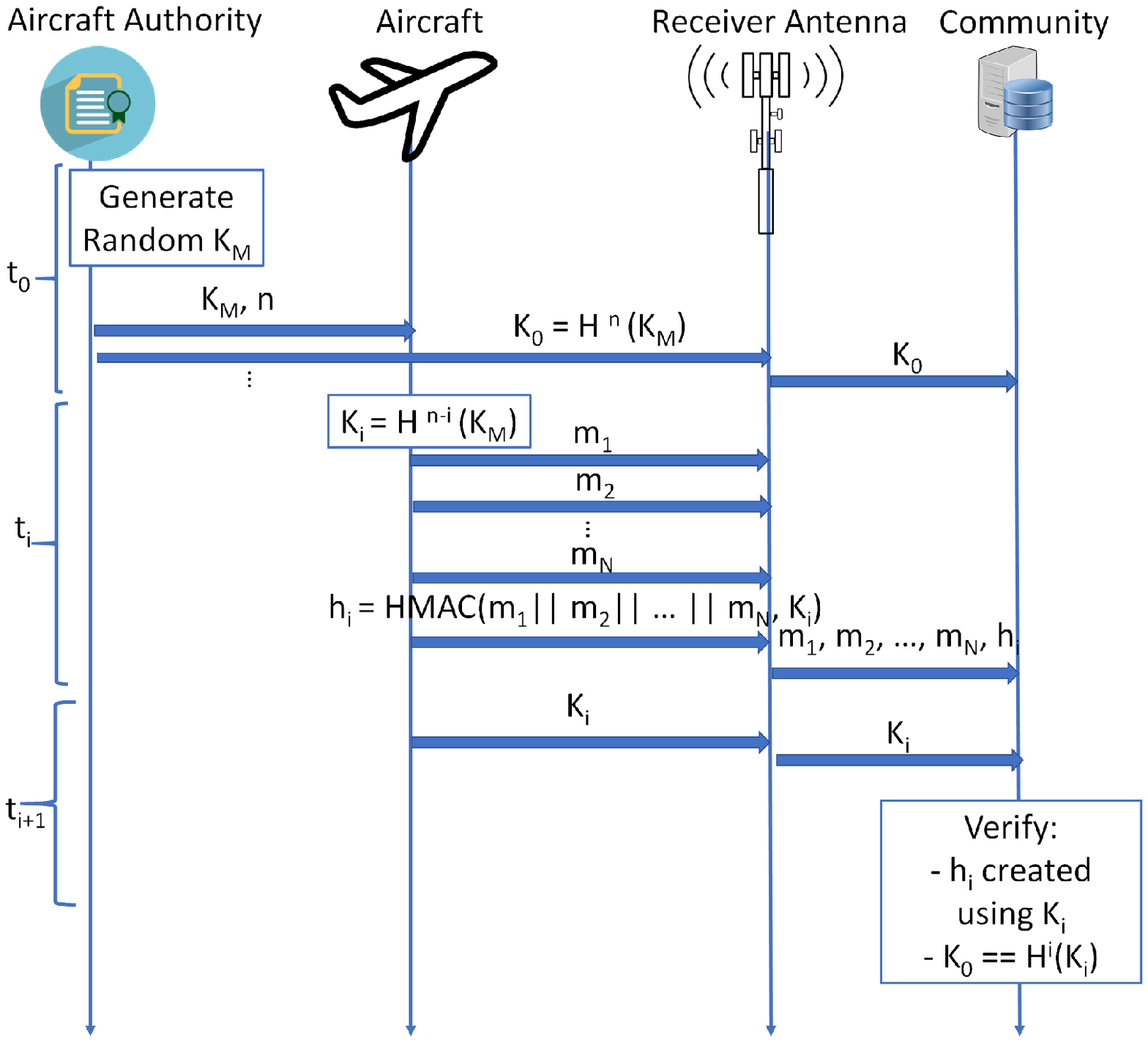}
	\caption{The \proto\ scheme.}
	\label{fig:proposal}
\end{figure}

From now on, we will assume that the Community Server (or, equivalently, the computational unit behind the receiver antennas) works in the \emph{Secured Mode}. 
The \proto\ scheme, depicted in Fig. \ref{fig:proposal}, can be divided in three distinct phases, that are the \emph{Setup Phase}, the \emph{Online Phase} and the \emph{Verification Phase}.

The steps performed in each of these phases are reported in the following.
\begin{itemize}
	\item \textbf{Setup Phase}. It is executed at the bootstrap of the flight by the Avionics Authority (i.e., a prominent authority, such as \ac{ICAO} or EUROCONTROL). Specifically, the Aircraft Authority equips the aircraft with the following elements:
	\begin{itemize}
		\item[-] a master key, $K_M$, that is a K bit key uniquely assigned to the particular aircraft for the duration of the flight;
		\item[-] an integer $n$, that is a large integer number representing the length of the hash chain.
	\end{itemize}
	Specifically, starting from the above two parameters, the root key $K_0$ of the aircraft is computed as:
	\begin{equation}
	\label{eq:root_key}
	K_0 = H( H( ...( H(K_M) ... ))) = H^n(K_M),
	\end{equation}
	where $H^n(K_M)$ refers to the execution of the hashing function $H$ on the input value $K_M$ for $n$ consecutive times.
	At the end of this phase, the Aircraft Authority makes public the following parameters:
	\begin{itemize}
		\item[-] the ICAO address of the flight, that is the unique identifier of the aircraft during the present flight;
		\item[-] the absolute value of $t_0$, that represents the boot-up time of the flight, i.e., the time in which the aircraft was equipped with the previous materials;
		\item[-] the root key $K_0$ of the aircraft, representing the key used by the aircraft to authenticate messages broadcast at the first useful slot.
	\end{itemize}
	All these parameters are shared through a publicly available server, that is supposed to be online at least for some time during the duration of the flight.
	
	\item \textbf{Online Phase}. Let us focus on the operation of the aircraft during the time-slot $t_i$, with $i>0$, and assume the aircraft actually delivers $N$ messages, $\left[ m_1, m_2, ..., m_n, ..., m_N \right]$, $N \geq 1$, during the time-slot $t_i$. 
	
	At some point in time, before the end of the slot, the aircraft computes the key for the current time-slot $t_i$, according to the following Eq. \ref{eq:key_t_i}:
	\begin{equation}
	\label{eq:key_t_i}
	K_i = H^{n-i}(K_M).
	\end{equation}
	
	The key $K_i$ is used by the aircraft to authenticate all the messages delivered during the time-slot $t_i$. To this aim, the aircraft generates a message digest $h_i$, by using a \ac{HMAC} function and the key $K_i$, as in the following Eq. \ref{eq:h_i}:
    \begin{equation}
	\label{eq:h_i}
	h_i = HMAC(m,K_i) = H ((K_i' \oplus opad) || H ( (K_i'\oplus ipad) || m )),
	\end{equation}
	
	where $K_i'$ is another secret key generated from the key $K_i$, the symbol $||$ refers to the concatenation operation, while $ipad$ and $opad$ are the well-known hexadecimal inner and outer constants, respectively \cite{rfc2104}.
	
	The digest $h_i$ is the element that allows for the verification of the pool of messages delivered within the time-slot $t_i$. Given that all the messages sent in that time-slot should be verified together, the aircraft delivers this message as the last of its pool, within the time-slot $t_i$. 
	
	Note that the receivers decode and store all the messages received by the aircraft. However, they still cannot validate them, given that they miss the information about the key $K_i$ used to generate the digest $h_i$. Thus, they temporarily store the messages in a buffer.
	
	\item \textbf{Verification Phase}. This final phase is dedicated to the verification of the messages delivered within the slot $t_i$, and it takes place at the beginning of the following slot, namely the $i+1$-th slot. 
	
	From the aircraft perspective, it consists in the delivery of a single-message, containing the key $K_i$ used by the aircraft to build the digest $h_i$ and to authenticate the messages sent in the time-slot $t_i$. The verification message is delivered by specifying a Payload Sub-Type field equal to 32. 
	
	When the ground stations receive the message, provided that they have received all the messages delivered by the aircraft in that time slot, they can verify the authenticity of all the messages received within the time-slot $t_i$. However, this is more likely to happen on the central server of the community controlling the particular receiver. Indeed, while some packets can be lost by some receivers hardly reached by the aircraft messages, it is very unlikely that a message is lost by all the receivers, since they enjoy a loose location correlation. This is further discussed in Sec. \ref{sec:benign}.
	
	This phase can be further divided in two sub-phases: the \emph{Normal Mode} and the \emph{Recovery Mode}.
	
	\paragraph{Normal Mode} In this sub-phase the verifier (either the single receiving sensor or the community server) checks the following conditions:
	\begin{itemize}
		\item It is possible to obtain the root key $K_0$ by hashing exactly $i$ times the key $K_i$, thus $K_0 = h^i(K_i)$;
		\item The received hash $h'_i$ is equal to the hash computed over all the messages received in the time-slot $t_i$, by using the key $K_i$; thus, $h'i = HMAC(m, K_i)$.
	\end{itemize}
    In this way, the set of community receivers can be confident that the messages were authenticated using the key $K_i$, and that the key could only be generated by the target aircraft, given that it is the only entity that could have generated it.
	Otherwise, if the second check is not verified for any of the active airplanes, it means that the target aircraft is under message injection attack. 
	Thus, the \emph{recovery mode} is triggered.
	
	\paragraph{Recovery Mode} The aim of this phase is to make an attempt to recover the set of legitimate messages. Specifically, the messages can be discarded, or an attempt to recover them can be performed as discussed below:
	\begin{itemize}
		\item Assume $M = N+J$ distinct messages have been received by the community server in the time-slot from a given aircraft, where $N$ is the number of legitimate messages and $J$ is the number of malicious messages. Note that $N$ is known to the Community Server, given that the number of messages between two consecutive \emph{Verification Key} messages is fixed. 
		
		The time within the time-slot bounds is further divided in a number $S$ of smaller sub-slots, each containing $L$ messages, 
		Within the sub-slot, the community server takes a decision based on majority voting. Thus, it selects the messages whose position is validated by the majority of the anchors. After applying the majority voting within all the slots, the community server ends up with a total of $T$ messages, with $T < M$.
		
		\item On the selected $T$ messages, assuming $N$ of these are legitimate messages, the community server tries all the possible combinations of messages, with the aim of finding the legitimate pool. Specifically, it evaluates all the possible groups of $N$ messages, checking that the digest computed through the verification key $k_i$ and the selected pool of messages is equal to the value $h_i$ previously delivered by the aircraft. 
		Thus, the maximum number of hash operations and comparisons required by the community server to find the correct sequence of messages is $\Delta = \binom{T}{N}$.
		If a valid pool is found, these are the authentic messages. Otherwise, no authentic messages are found for the time-slot $t_i$ and the messages are discarded.
	\end{itemize}
\end{itemize}

It is worth noting that the strategy implemented in the \emph{Verification Phase} of \proto\ is indeed effective against an adversary that injects fake position messages of the target aircraft, being this position totally different from the real one. In addition, realistic adversaries emit their messages with a \ac{SDR} that is located at the ground-level. Being the \ac{ADS-B} technology very sensitive to the presence of obstacles \cite{Schafer2014}, the expected number of receivers for the fake messages is lower than the legitimate ones, that are emitted at greater altitudes, with a reduced probability to find obstacles and thus higher chances to be received by a greater pool of anchors. 
Otherwise, if the attacker is able to force the reception of the fake message by many anchors (i.e., by using ADS-B equipped drones), the maximum benefit it can expect is to cause a \ac{DoS} on the system, given that none of the authentic messages will be accepted. 

\section{Performance Assessment}
\label{sec:performance}

\subsection{Benign Scenario}
\label{sec:benign}

In this section we evaluate the performances of \proto\ in a benign scenario, with the aim of gaining more insights on its bandwidth and computational requirements in standard operational conditions.

\begin{figure}[htbp]
	\centering
	\includegraphics[width=.6\columnwidth]{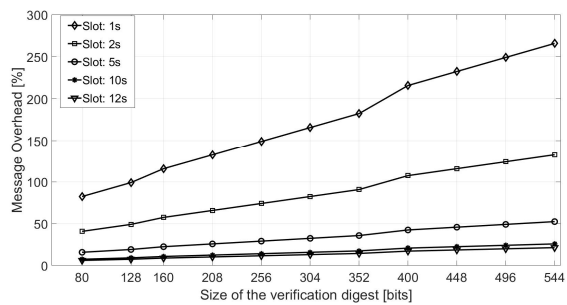}
	\caption{Overhead derived by the adoption of \proto, by considering different lengths of the verification digest and different duration of the time slot.}
	\label{fig:hash_slot}
\end{figure}

In Fig. \ref{fig:hash_slot} we illustrate the bandwidth overhead of \proto\ with respect to the size of the verification digest and the duration of the time-slot, by assuming a fixed 128-bit verification key.
As the length of the verification digest increases, both the security provided to the messages and the message overhead increase, given that more messages need to be delivered over the radio interface. At the same time, the overhead lowers as the time-slot duration increases, given that more messages are authenticated using the same digest. It is worth noting that the same considerations are valid if we increase the key size, while fixing a specific digest size. 
As the security of \proto\ lies in the size of both the verification key and verification digest, a compromise between the bandwidth overhead and the security level is required. In general, assuming both a key length and a verification digest of 128 bits, and assuming to fix a 2 seconds long time-slot, the bandwidth overhead introduced by \proto\ is 47.58\%, that is we use roughly the 50\% of the messages to authenticate the batch of messages sent within the time-slot.  Note that this overhead can be considered both as included in the actual throughput of a peer-to-peer communication, or added as an additional overhead to the actual rate of the \ac{ADS-B} technology. In the second case, this leads to an increase of the maximum packets rate from 6.2 to 9.14 packets/sec. Given that the ICAO standard envisions situations in which the maximum recommended rate can be exceeded, this is not a violation of the standard.

\subsection{Comparison and Discussion}
\label{sec:comparison}

Still assuming a benign scenario, in this section we compare the performance of \proto\ with closely related work, by considering the size of the cryptography materials (keys and digest size), the bandwidth overhead, and the compliance to the standard of all the solutions. The main results have been reported in Tab. \ref{tab:comparison}.
\begin{table*}[htbp]
	\caption{Comparison with security approaches published in \cite{Berthier2017}, \cite{Yang2017_globecom} and \cite{Yang2017}.}
	\label{tab:comparison}
	\centering
	\begin{tabular}{|p{2.2cm}| p{0.9cm}| p{1.1cm}| p{1.9cm} | p{1.5cm}| p{1.8cm}| p{1.6cm}|}
		\hline
		\textbf{Scheme} & \textbf{Key Size [bits]} & \textbf{Digest Size [bits]} & \textbf{Crypto Parameters Soundness} &\textbf{Slot Duration} & \textbf{Std. compliance} & \textbf{Overhead [\%]}\\
		\hline
		\textbf{\proto} & 128 & 128 & \cmark & 2 s & \cmark & 47.58 \\
		\hline
		\textbf{SAT \cite{Berthier2017}} & 128  & 16 & \xmark & 5 s & \xmark & 22.9 \\
		\hline
		\textbf{LHCSAS \cite{Yang2017_globecom}} & 80 & 128 & \cmark & 1 msg. & \xmark & 500 \\
		\hline
		\textbf{HIBS \cite{Yang2017}} & N/A & 1,024 & \cmark & - & \cmark & 2,200 \\
		\hline
	\end{tabular}
\end{table*}

SAT \cite{Berthier2017} is based on the TESLA authentication primitive, but it is not standard-compliant. In fact, its authors include the digest of each message within the related ADS-B packet just before the PI field, thus modifying the message length imposed by the standard.  
In addition, every 30 seconds the protocol recommends the broadcast of a certificate including a key of 128 bits, signed through a public key of 512 bits and the \ac{ECDSA} technique. Finally, independently from the particular hashing algorithm used, SAT constrains the digest to be 16-bits long, hence jeopardizing the security of the proposed scheme. 
Assuming that the certificate is generated through the well-known \emph{openssl} tool, it results in a minimum overhead of 22.9\%. 

LHCSAS \cite{Yang2017_globecom} still breaks the compatibility with the standard: in fact, it modifies the mandatory \emph{subType} field, replacing it with cryptography data. In addition, the aircraft delivers cryptography elements for each message, thus generating 5 additional packets for every ADS-B message.

HIBS \cite{Yang2017} adopts robust cryptography properties. In fact, packets are authenticated through a digest of 1024 bits. However, a digest of such a size is generated for each packet, resulting in an enormous bandwidth overhead. By assuming to work with the extended version of the scheme and maintaining the size of the message imposed by the standard, 22 additional messages are necessary for each payload to be authenticated, resulting in a dramatic bandwidth increase of 2200\%.
Instead, \proto\ integrates authentication services based on symmetric encryption within the ADS-B payload in a standard-compliant fashion. The resulting overhead, as per what discussed in the previous subsection, is 47.58/\%. 
This slight higher overhead, however, is compensated by the enhanced security level provided to the \ac{ADS-B} technology.

\begin{figure}[htbp]
	\centering
	\includegraphics[width=.6\columnwidth]{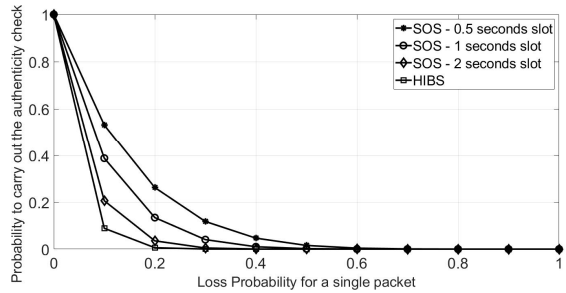}
	\caption{Loss probability for a single packet.
	}
	\label{fig:perf_comparison}
\end{figure}

\proto, as all the other solutions that do require packet fragmentation, is vulnerable to packet loss. In fact, if a single packet delivered within the whole time-slot is not received by any of the ground receivers, all the packets within the same slot cannot be verified \cite{sciancalepore_sac2019}. 
In general, the deployment of a large number of antennas improves the probability that at least one of them receives a packet. Even if packet loss is theoretically always possible, it is worth noting that an high level of packet loss disrupts also the correct functioning of the other computing solutions discussed above. 
Neglecting not standard-compliant approaches and assuming different values of the slot duration of \proto, Fig. \ref{fig:perf_comparison} evaluates the probability to successfully receive all the elements necessary to carry out the authenticity check, both with \proto\ and with \cite{Yang2017}, with an increasing loss probability on the overall system.

\proto\ cannot verify the authenticity of a single packet if at least a message transmitted in the time-slot of duration $2$ seconds is lost. Assuming a default transmission rate of 6 packet/s, the loss could occur in any of the 12 messages sent within the time-slot, or in the 3 messages delivered in the next slot and containing the verification key. Thus, there would be at least a single packet loss in 15 messages. However, HIBS requires the correct reception of 23 messages to evaluate the authenticity of the information. Thus, the packet loss would be more disruptive in the proposal by \cite{Yang2017} than in the \proto\ scheme. This is still true also in case packet losses happen in burst, given that \proto\ could provide, under reasonable assumptions, intermittent connectivity with the community server.
\subsection{Scenario with a malicious adversary}
\label{sec:malicious}

In this section we evaluate the performance of \proto\ and the contribution in \cite{Yang2017} in the presence of a malicious active adversary.

\begin{figure}[htbp]
	\centering
	\includegraphics[width=.6\columnwidth]{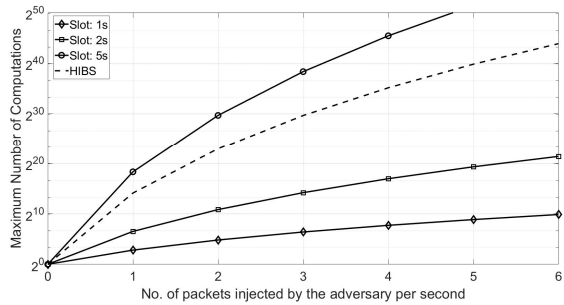}
	\caption{Number of required computations by \proto\ and \cite{Yang2017} on the community servers, under the hypothesis of attack by a malicious adversary.
	}
	\label{fig:comparisons}
\end{figure}

During a given time-slot, the adversary injects fake packets in the wireless communication medium, with the aim of confusing the receivers about the current position occupied by the legitimate aircraft. 
In case of an active attack, the second check performed in the verification phase of \proto\ fails. Specifically, the digest computed over all the messages received by the community server from the target aircraft, through the key $K_i$ of the current slot $t_i$, will not be equal to the verification digest $h_i$.
In this situation, the community server triggers the \emph{Recovery Mode}. Thus, it first adopts an approach based on majority voting, by discarding messages claiming a given position but received by the minority of the anchors within a given sub-slot.
On the remaining messages, the community server checks for the pool of messages that verifies the authenticity check. This is indeed possible thanks to the fixed number of packets between two consecutive digests. The performance of \proto\ and HIBS in this situation are showed in Fig. \ref{fig:comparisons}, assuming the maximum transmission rate by the legitimate aircraft of 6 pkts/sec.

Focusing on the performance of \proto, the figure shows that the shorter the time-slot, the less the maximum number of operations that are required on the community server's side.
Assuming a short duration of the time-slot, i.e., 1 second, and that the adversary injects malicious packets with a rate of 6 pkts/sec, the number of operations required by the community server would be equal to about 924, indeed a tolerable amount of HMAC for the community server. Of course, the higher the rate of transmission by the adversary, the higher the computational overhead by the aircraft.
This becomes an issue by assuming an higher duration of the time-slot, resulting in an unmanageable maximum number of comparisons when the duration of the slot is equal or higher to 5s. The same issue emerges with the usage of the HIBS protocol. Assuming the transmission rate of 6 packets/sec, HIBS requires almost 4 seconds to deliver a single information packet, along with all the security material. If the attacker injects packets at a rate of 6 pkts/sec, this would result in more than $2^{42}$ maximum computations, indeed a very resource-consuming task.
By looking at results showed in Sec. \ref{sec:benign}, the time-slot duration of the \proto\ protocol must be carefully selected in order to trade-off between the bandwidth overhead and the number of comparisons to deal with in case of attack. 
For instance., by assuming to work with a time-slot duration of 2 seconds, in case the adversary injects 6 packets, the community-server will require about a maximum number of $2^{21}$ hashes and comparisons to find the authentic pool of messages. According to latest measurement with dedicated hardware (\url{https://gist.github.com/epixoip/a83d38f412b4737e99bbef804a270c40}), about 2.25 seconds are necessary to find the legitimate pool of messages. Other measurements with non-dedicated hardware can be obtained through public data (\url{https://en.bitcoin.it/wiki/Non-specialized_hardware_comparison}).

\section{Conclusions and Future Work}
\label{sec:conclusion}

Inspired by its mandatory adoption on board of all commercial aircraft by the 2020, and pressed by its anticipated adoption by major airlines (e.g., Qatar Airways, American Airlines and British Airways), in this paper we proposed \proto, a lightweight and standard-compliant framework designed to guarantee the authenticity of the communications in the ADS-B technology.
The framework integrates the $\mu$-TESLA protocol in ADS-B communications, allowing to batch-verify all the messages originated by an airplane in a given time-slot. In addition, the framework leverages a majority voting filtering stage in the message reception phase and it is suitable for deployment on community-oriented services, as the emerging OpenSky-Network community. 
Moreover, it is resilient to active attacks attempting to poisoning the message authentication process. Finally, comparisons with state of the art solutions do show that \proto\ is the winning solution in terms of provided security and achieved performance.

Future research activities include refining the packet loss hypothesis (studying packet burst loss model) and the implementation of the proposed framework using commercial Software Defined Radios.

\bibliographystyle{splcns}
\bibliography{adsbsecurity}

\end{document}